\documentclass[%
 reprint,
 amsmath, amssymb,
aps,
prstab,
longbibliography,
]{revtex4-2}

\usepackage{color}
\usepackage{graphicx}% Include figure files
\usepackage{dcolumn}% Align table columns on decimal point
\usepackage{bm}% bold math
\usepackage[utf8]{inputenc}      % switch to utf8
\usepackage{textcomp} % support Greek letter
\usepackage{multirow}
\usepackage[mathlines]{lineno}% Enable numbering of text and display math
\begin{document}

\preprint{APS/123-QED}

\title{Beam halo from Touschek scattering in the KEK Accelerator Test Facility}
%\thanks{}
\author{R. Yang$^{1,2}$}\email{renjun.yang@kek.jp}
\author{A. Aryshev$^{1,2}$, P. Bambade$^{3}$, M. Bergamaschi$^{4}$, \\ K. Kubo$^{1,2}$, T. Naito$^{1,2}$, N. Terunuma$^{1,2}$, S. Wallon$^{3}$}
\author{J. Zhang$^{5}$}
\affiliation{
    $^{1}$High Energy Accelerator Research Organization,  Tsukuba 305-0801, Japan\\
    $^{2}$School of High Energy Accelerator Science, SOKENDAI, Tsukuba 305-0801, Japan\\
    $^{3}$IJCLab, CNRS/IN2P3, Universit$\acute{e}$ Paris-Saclay, Orsay 91898, France\\
    $^{4}$European Organization for Nuclear Research, Geneva CH-1211, Switzerland \\
    $^{5}$Deutsches Elektronen-Synchrotron, Hamburg D-22607, Germany
    }%
%
%\collaboration{MUSO Collaboration} %\noaffiliation
%\affiliation{
%\affiliation{
%\affiliation{%

\date{\today}% It is always \today, today,

\begin{abstract}
Beam halo is one of the crucial issues limiting the machine performance and causing radioactivation in high-intensity accelerators. A clear picture of beam-halo formation is of great importance for successful suppression of the undesired beam loss. We present numerical and experimental studies of transverse and longitudinal halos in the KEK Accelerator Test Facility. The observed general consistency between predictions and observations in various conditions indicates that the Touschek scattering is the dominant mechanism forming the horizontal and momentum halos.  

\begin{description}
\item[DOI]

\end{description}
\end{abstract}

\pacs{11}% PACS, the Physics and Astronomy
\maketitle

%\tableofcontents
\section{\label{sec:intro}INTRODUCTION}
% why study halo? 
% accel.--> beam loss -> coll. -> halo meas./model -> simu./meas. (??)
Beam halo is one of the most critical issues limiting the performance and potentially causing component activation for high-intensity accelerators, especially high-energy colliders at the luminosity frontier. The formation of beam halos is generally complicated and associated with various collective effects, nonlinearity, optics errors, beam-beam interaction, secondary emission, and so on~\cite{hirata1992nongaussian, chen1994simu, helmut2000trans, burkhardt2008halo, gluckstern1994ana, allen2002beam, oide1994anomalous, ikegami1999, hofmann2016prl, wittenburg2009beam}. To mitigate the undesired background induced by halo particle loss, a robust collimation system is an indispensable part of a high-intensity accelerator. To accurately estimate collimation efficiencies and residual backgrounds, good knowledge of the primary driving mechanisms and adequate modeling is of vital importance. 
%Besides, a clear picture of halo formation can help to predict background in some sensitive region, e.g., the injection point and interaction point.
% To accomplish this mission, both simulations with various physical processes and observations employing a powerful halo monitor with sufficient dynamic range are essential.

% what's ATF? parameters. 
% why study halo at ATF? progresses, unresolved questions, why un-resolve?
 The Accelerator Test Facility~(ATF) at KEK was initially constructed to demonstrate the feasibility of producing low-emittance beams required at a future linear collider and supply high-quality beams for the R{\&}D activities on beam dynamics, instrumentation and control technology which will be needed at such facilities~\cite{PhysRevLett.88.194801, PhysRevLett.92.054802, white2014experimental, atf2rep2020}. It consists of a 1.3~GeV injector, a race-track style damping ring~(DR) and an extended extraction line.  The major parameters of the DR are listed in Table~\ref{tab:atfpara}. ATF provides an excellent opportunity to investigate halo formation towards future colliders. Compared with a high-energy collider with typically high local chromaticity in the interaction region and strong nonlinear beam-beam interaction effects, the primary mechanisms driving particles into the halo region in a GeV-scale electron storage ring are more straightforward, involving mainly beam-gas scattering~(BGS), Touschek scattering and nonlinearity~\cite{wittenburg2009beam, boscolo2012monte, breunlin2016tous, andrii2021bgcoll}. However, the verification of such plausible driven mechanisms from direct observations has been rarely reported. Prior beam halo studies in ATF have concentrated on the development of high-dynamic-range halo monitors and analytical evaluations based upon Campbell's theorem~\cite{liu2016vacuum, naito2016beam, dou2014analytical}. Quantitative measurements using diamond sensor based detector did show that the vertical halo is dominated by the elastic BGS process~\cite{yang2018evaluation}. However, the horizontal halos were much larger than predicted by BGS, and could not be fully explained. Given the relatively large non-zero horizontal dispersion, it was suspected that they were arisen from  an inelastic scattering process such as Touschek scattering, but a careful check of this hypothesis could not be pursued due to the lack of an adequate numerical simulation with all the necessary physical processes and a powerful monitor for both transverse and longitudinal halos.

In this article, numerical simulations of halo generation including a more complete set of scattering processes in the presence of realistic machine parameters are presented, followed by experimental observations of transverse and longitudinal halos employing a combined YAG/OTR screen monitor. The reasonable agreement between these new simulations and the measurements demonstrates the Touschek effect's leading role in the formation of horizontal and momentum halos.
\begin{table} [h]
\caption{\label{tab:atfpara} ATF main parameters~\cite{PhysRevLett.88.194801, PhysRevLett.92.054802}.}
\begin{ruledtabular}
		\begin{tabular}{lc}
		Beam energy [GeV] &	1.3 \\
		% \midrule
		Circumference [m] & 138.6 \\
		Bunch charge [nC] &   0.16-1.6\\
		Vertical emittance [pm] 	& $>$4	\\
		Horizontal emittance [nm]		&	1.2	\\
		Energy spread [\%]	&	0.056 (0.08)\footnote{For a bunch charge of 1.6~nC.}	\\
		Bunch length [mm]	&	5.3	(7)$^\text{a}$\\			
		Number of bunches & 1-20 \\
		Repetition rate [Hz]&  3.12 \\	
		\end{tabular}	
\end{ruledtabular}
\end{table}

\section{\label{sec:simu} Simulations}
The simulation which is used includes three main parts: mimicking of realistic beam parameters, generation of halo particles from stochastic processes, and particle tracking. The halo generator was developed based on SAD~\cite{SAD}. It initially included only the BGS process~\cite{yang2017numerical, yang2018evaluation}, and has now been expanded to treat also Touschek scattering. 

Unlike the BGS process, Touschek scattering closely depend on the particle density. To reproduce the operational emittances, vertical dispersion and $xy$ coupling are deliberately introduced through local-dispersion bumps and rotations of quadrupoles in the straight sections, respectively~\cite{yang2020bw}. Chromaticity is controlled by the corresponding sextupole families, while the beta-beat and horizontal-dispersion errors are ignored. For high-intensity beams, equilibrium emittances can also be significantly diluted due to the intra-beam scattering~(IBS) process and are numerically evaluated with the beam-envelope method~\cite{ohmi1994beam, kubo2001ibs}.
%, as shown in Fig.~\ref{fig:emit_nb}.

% should we expand F(tm), i.e., Eq. (3-4) ????
Both elastic and inelastic scatterings between particles and nuclei of the residual gas have been included, as described in Ref.~\cite{yang2017numerical, yang2018evaluation}. For the sake of simplicity, a uniform gas pressure with CO as major gas component was assumed. For the Touschek effect, the theory established by Piwinski~\cite{piwinski1999touschek} has been employed to evaluate momentum transformations and the relevant two-dimensional~(2D) particle distributions and variations of beam envelopes in the presence of non-zero dispersion. For a bunched beam with Gaussian phase-space distributions, the rate of Coulomb scattering leading to a longitudinal momentum perturbation of [$\Delta\delta_1$, $\Delta\delta_2$] can be expressed as
\begin{equation}\label{eq:tousRate_delta}
    R(\Delta\delta_{1}, \Delta\delta_{2}) = R(\Delta\delta_1, \infty) - R(\Delta\delta_2, \infty)
\end{equation}
with 
\begin{equation}\label{eq:tousRate_piwinski41}
	R(\Delta\delta, \infty) = \frac{r_e^2cN_p^2}{8\pi\gamma^2\sigma_z\sqrt{\sigma_x^2\sigma_y^2-\sigma_\delta^4\eta_x^2\eta_y^2}\tau_m}F(\tau_m)
\end{equation}
where $r_e$ is the classical electron radius, $c$ the velocity of light in vacuum, $N_p$ the bunch population, $\gamma$ the Lorentz factor, $\eta_{x,y}$ the dispersion, $\sigma_{x, y}$ the transverse beam size, $\sigma_\delta$ the energy spread, $\sigma_z$ the bunch length and  ${\tau_m=\beta^2\Delta\delta^2}$, with $\beta$ the relative velocity of beam. $F(\tau_m)$ is an integration over the whole beam column, as
\begin{equation}
    \begin{aligned}
    F(\tau_m) = & \sqrt{\pi(B_1^2-B_2^2)}\tau_m\int_{\tau_m}^{\infty}\Big[ \Big( 2+ \frac{1}{\tau} \Big)^2 \Big( \frac{\tau/\tau_m}{1+\tau} - 1\Big) \\
     & +1 - \frac{\sqrt{1+\tau}}{\sqrt{\tau/\tau_m}} - \frac{1}{2\tau}\Big( 4 + \frac{1}{\tau} \Big)\log\Big( \frac{\tau/\tau_m}{1+\tau} \Big) \Big] \\ 
     & e^{-B_1\tau}I_0(B_2\tau)\frac{\sqrt{\tau}d\tau}{\sqrt{1+\tau}}
    \end{aligned}
\end{equation}
with 
\begin{equation}
	\begin{aligned}
		B_1 & = \frac{\beta_x^2}{2\beta^2\gamma^2\sigma_{x\beta}^2}(1 - \frac{\sigma_h^2\tilde{\eta}_x^2}{\sigma_{x\beta}^2}) + \frac{\beta_y^2}{2\beta^2\gamma^2\sigma_{y\beta}^2}(1 - \frac{\sigma_h^2\tilde{\eta}_y^2}{\sigma_{y\beta}^2}) \\
		B_2^2 & = B_1^2 - \frac{\beta_x^2\beta_y^2\sigma_h^2}{\beta^4\gamma^4\sigma_{x\beta}^4\sigma_{y\beta}^4\sigma_{\delta}^2}(\sigma_x^2\sigma_y^2 - \sigma_{\delta}^4\eta_x^2\eta_y^2) \\
		\frac{1}{\sigma_h^2} & = \frac{1}{\sigma_{\delta}^2} + \frac{\eta_x^2 + \tilde{\eta}_x^2}{\sigma_{x\beta}^2} + \frac{\eta_y^2 + \tilde{\eta}_y^2}{\sigma_{y\beta}^2}
	\end{aligned}
\end{equation}
where $I_0$ is the modified Bessel function, $\beta_{x,y}$ the beta function, $\sigma_{x\beta, y\beta}$ the betatron beam sizes, ${\tilde{\eta}_{x,y}=\alpha_{x,y}\eta_{x,y}+\beta_{x,y}\eta'_{x,y}}$ and $\tilde{\sigma}_{x,y}^2=\sigma_{x,y}^2+\sigma_{\delta}^2\tilde{\eta}_{x,y}^2$.
The induced momentum perturbations~($\pm\Delta\delta$) are applied to two core particles, generated randomly based on the equilibrium beam matrices. The local scattering rate is determined taking into account the distance to the closest upstream element and then integrated to give the total scattering rate. These derivations assume a significant longitudinal momentum change and special attention should be paid to the choice of $\Delta\delta$. %Applying Eq.~(\ref{eq:tousRate_delta}-\ref{eq:tousRate_piwinski41}) for a small momentum transfer, the momentum tail near beam core might be underestimated~(see Sec.~\ref{sec:meas}).
Here, the minimum momentum change is set to $3\sigma_{\delta}$, while the maximum perturbation should be larger than the momentum acceptance~($\sim$1.2\%) since some large off-momentum particles might survive for a few turns and slip into the adjacent RF bucket. For simplicity, the transverse heating due to momentum transfers is not included, and therefore, the transverse diffusion takes place only in the dispersive regions.
\begin{figure} %[h]
    \centering
    \includegraphics[width=0.95\linewidth]{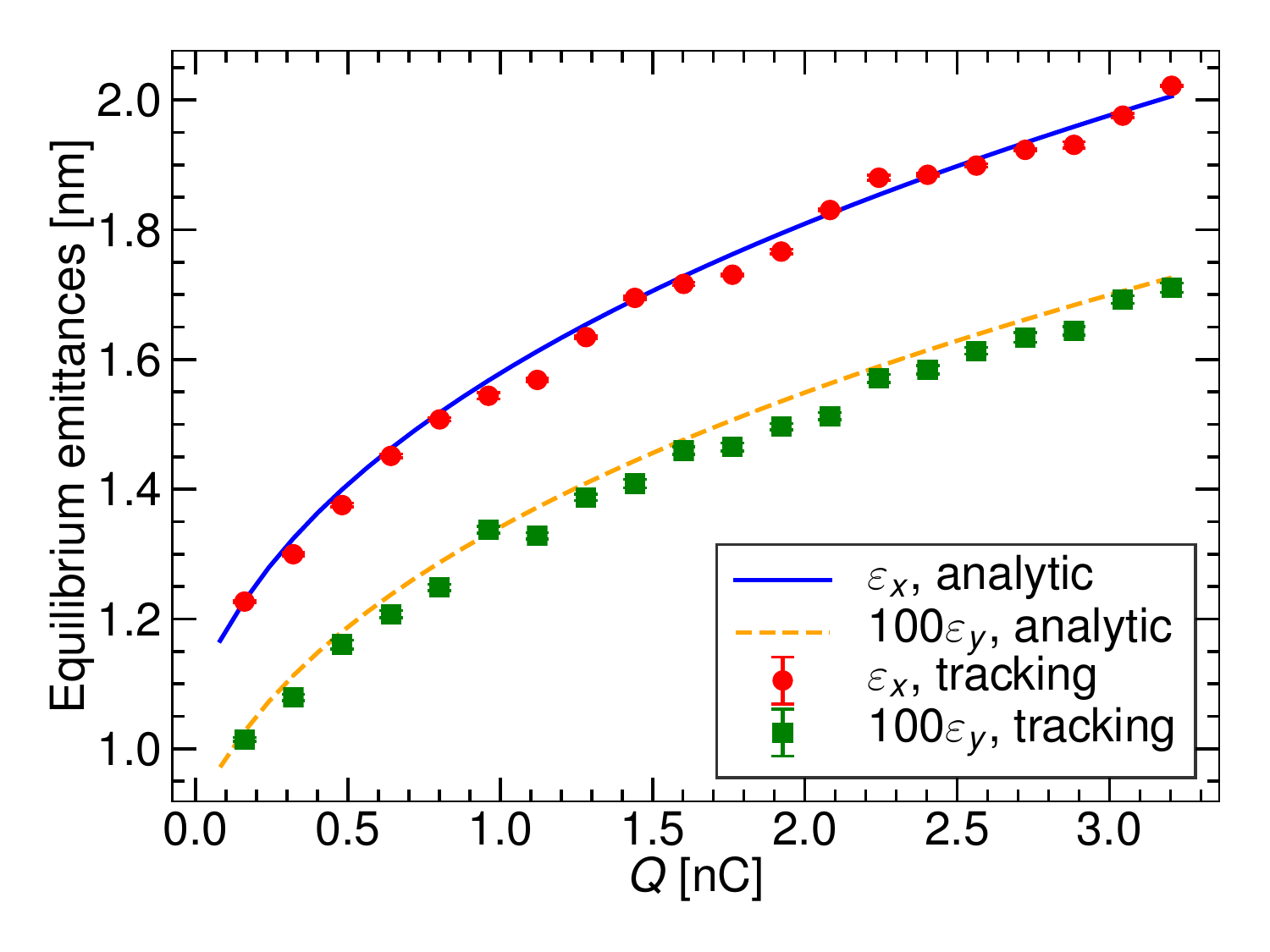}
    \caption{Equilibrium emittances evaluated by numerical calculation in the presence of synchrotron radiation and IBS, and tracking with one-turn damping and excitation matrices.}
    \label{fig:emit_nb}
\end{figure}

Scattered particles are tracked element-by-element using the default symplectic tracking routine of SAD. Radiation damping and diffusion from quantum excitation and IBS are applied in a turn-by-turn manner utilizing the corresponding matrices. In the ATF damping ring, transverse emittances evaluated analytically are in agreement with the tracking results, as shown in Fig.~\ref{fig:emit_nb}. The scattered particles are typically generated and tracked for two damping times to reach equilibrium. Thanks to the randomness of quantum and IBS fluctuations, one can accumulate halo particles over the last few turns for sufficient statistics with a reasonable computing time. The complete distributions are combinations of scattered particles (more than 1$\times10^8$) and core particles obtained through tracking in parallel.

\section{\label{sec:meas} Measurements}
\subsection{\label{subsec:moni} Experimental setup}
\begin{figure} %[h]
    \centering
    \includegraphics[width=\linewidth]{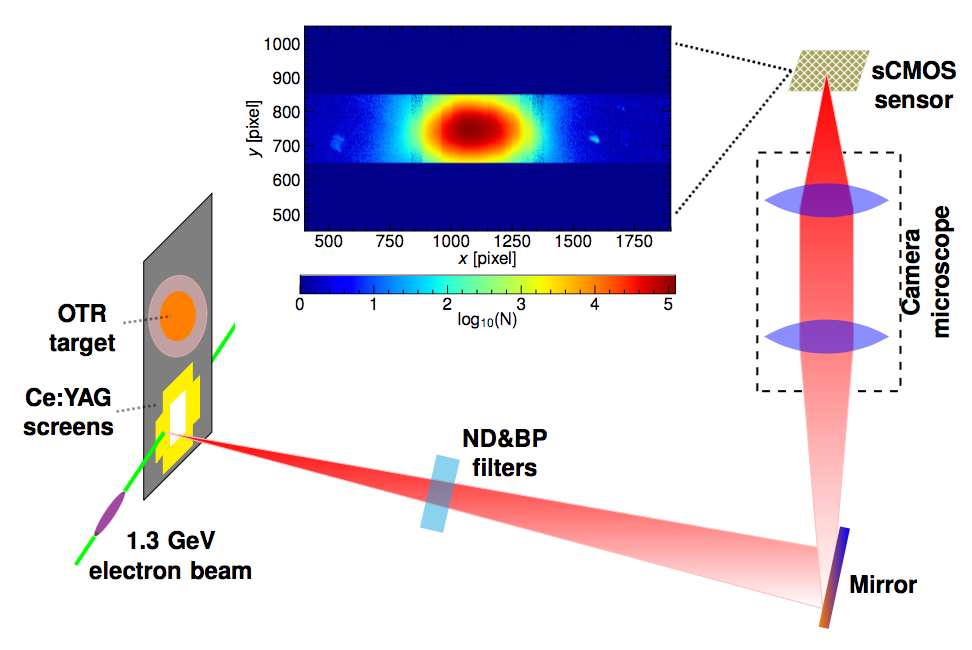}
    \caption{Schematic of the YAG/OTR monitor including a 2D core-halo image obtained through horizontal scanning of the left and right YAG pads. The glare in the halo region result from dust on the screen.}
    \label{fig:yag_image}
\end{figure}

For fast halo diagnostics, a combined yttrium aluminium garnet~(YAG)/optical transition radiation~(OTR) monitor has been developed~\cite{yang2018ibic}. Four 0.5~mol\%~Ce:YAG screens with a central rectangular opening are for visualizing core and halo profiles, and an OTR target provides supplementary visualisation of the dense beam core, for which the YAG image is deformed owing to scintillation saturation, as depicted in Fig.~\ref{fig:yag_image}. 
The screens are placed on a holder actuated by an automatic manipulator. For halo imaging, one must adjust the YAG pads to allow core particles to pass through the central opening. The YAG and OTR screens are at 45$^\circ$ and 67.5$^\circ$ to the beam trajectory, respectively, to collect light with a common optical system comprising filters, a microscope lens and a 16-bit complementary metal-oxide-semiconductor~(CMOS) camera~\cite{pcoedge42l}. 

To avoid blooming effects, halo distributions are measured through one-dimensional scans of the YAG pads. After taking a picture of the halo far from the beam core, the YAG pads are moved toward the beam core step-by-step and images are captured with light attenuation at each step. Then, sliced halo images at different distances to the center are cut out by trimming the parts near the inner edge of the YAG, and overlapped with preceding images. Combining the core and the sliced halo images of the two sides, a complete core-halo distribution is eventually obtained, as illustrated in Fig.~\ref{fig:yag_image}. Owing to this scanning procedure, vertical and horizontal profiles have to be captured individually. Using solely the YAG screens, a dynamic range of about $1\times10^5$, limited by the photon-yield efficiency of the scintillator, background noise and scintillation saturation, has been demonstrated~\cite{ryangthesis}. 

\begin{figure} %[h]
    \centering
    \includegraphics[width=0.95\linewidth]{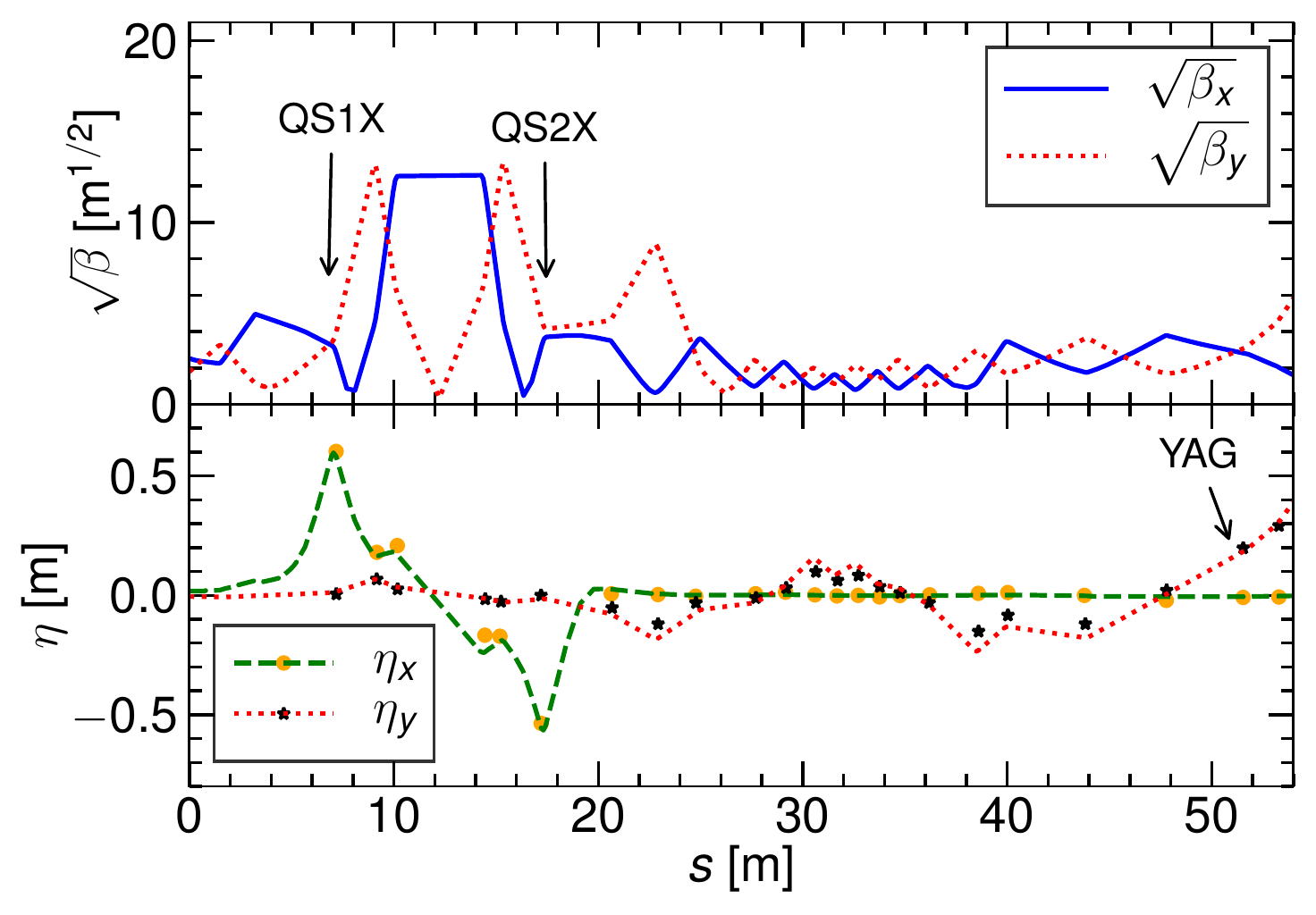}
    \caption{Beta-function and dispersion along the extraction line. The colored marks represent the experimental data and the lines denote the model optics. QS1X and QS2X are the skew-quadrupoles for vertical-dispersion control.}
    \label{fig:etay_bump}
\end{figure}

%\begin{figure} %[h]
%    \centering
%    \includegraphics[width=0.75\linewidth]{Reconstructed_image_hor_log_20171208}
%    \caption{A 2D beam profile obtained through horizontal scanning of left and right YAG pads. The glares in the halo region is suspected to be owing to dusts on the screen.}
%    \label{fig:yag_image}
%\end{figure}
For visualizing the momentum halo, the YAG/OTR monitor has been placed downstream of a dogleg inflector in the extraction line. A large vertical dispersion~($>$200~mm) can be easily bumped using two skew-quadrupoles located in the dogleg where the horizontal dispersion reaches its largest absolute values but with opposite signs~\cite{woodley2013atf2}, as shown in Fig.~\ref{fig:etay_bump}. 
%The vertical dispersion can be controlled by two skew-quadrupoles located in where with largest horizontal dispersion but opposite in sign~\cite{woodley2013atf2}. For the visualization of momentum halo, the vertical dispersion at the monitor could easily be increased to about 200~mm, as shown in Fig.~\ref{fig:etay_bump}. 

\subsection{\label{subsec:meas-trans} Transverse and longitudinal halos}
\begin{figure*} 
    	\centering
    	\includegraphics[width=0.49\linewidth]{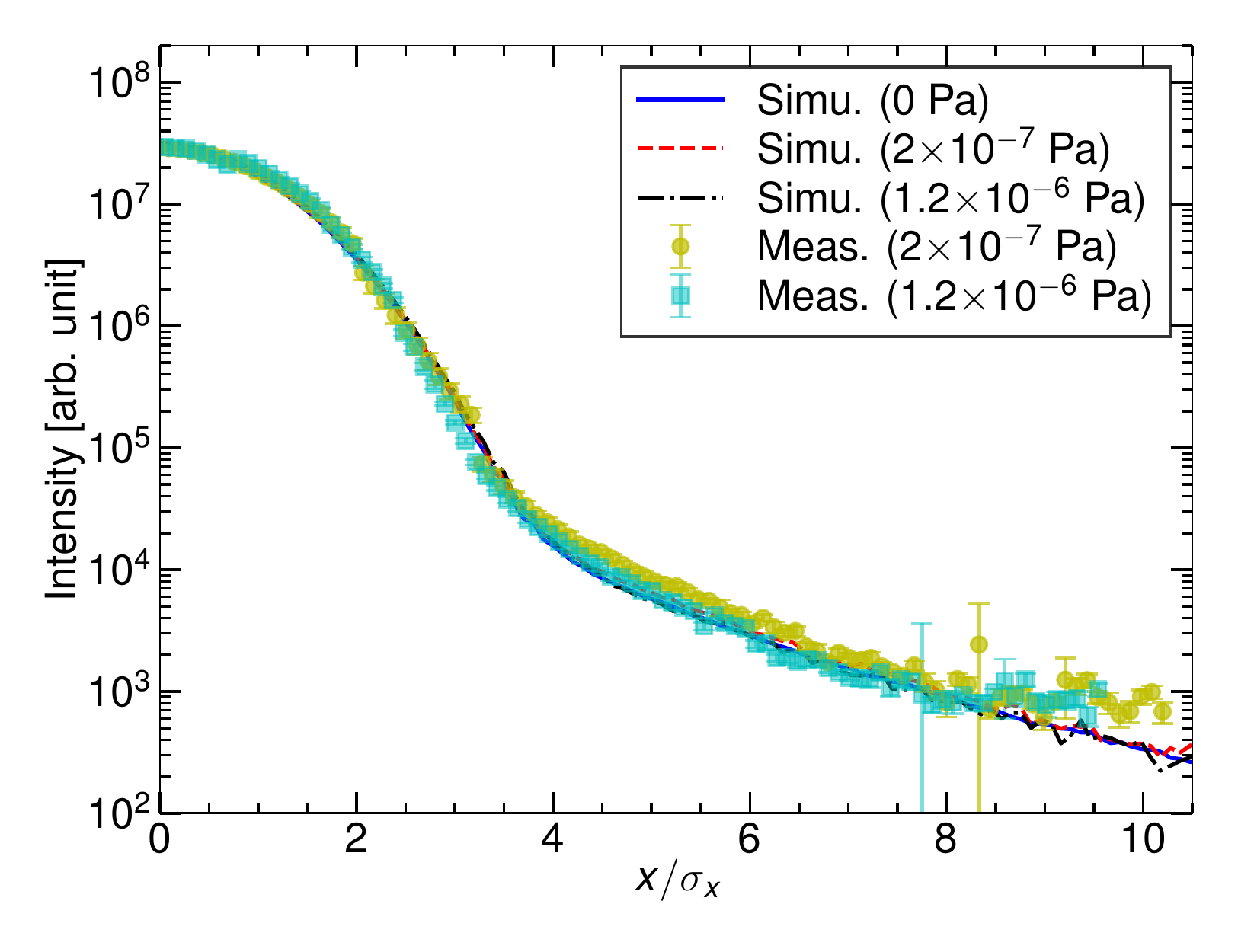}
    	\put(-200, 50){(a)}%\\    	
    	\includegraphics[width=0.49\linewidth]{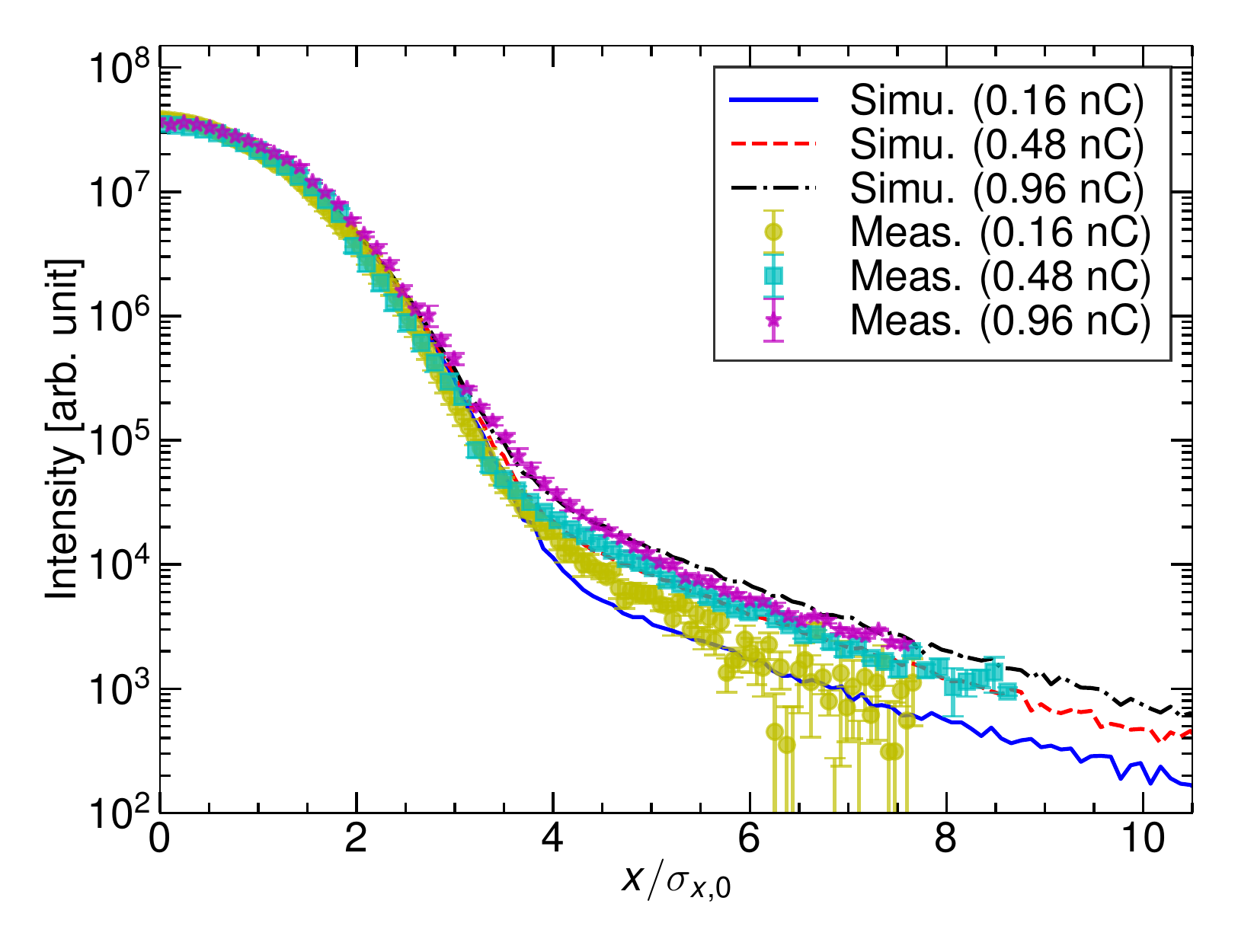}
    	\put(-200, 50){(b)}	\\
	    \includegraphics[width=0.49\linewidth]{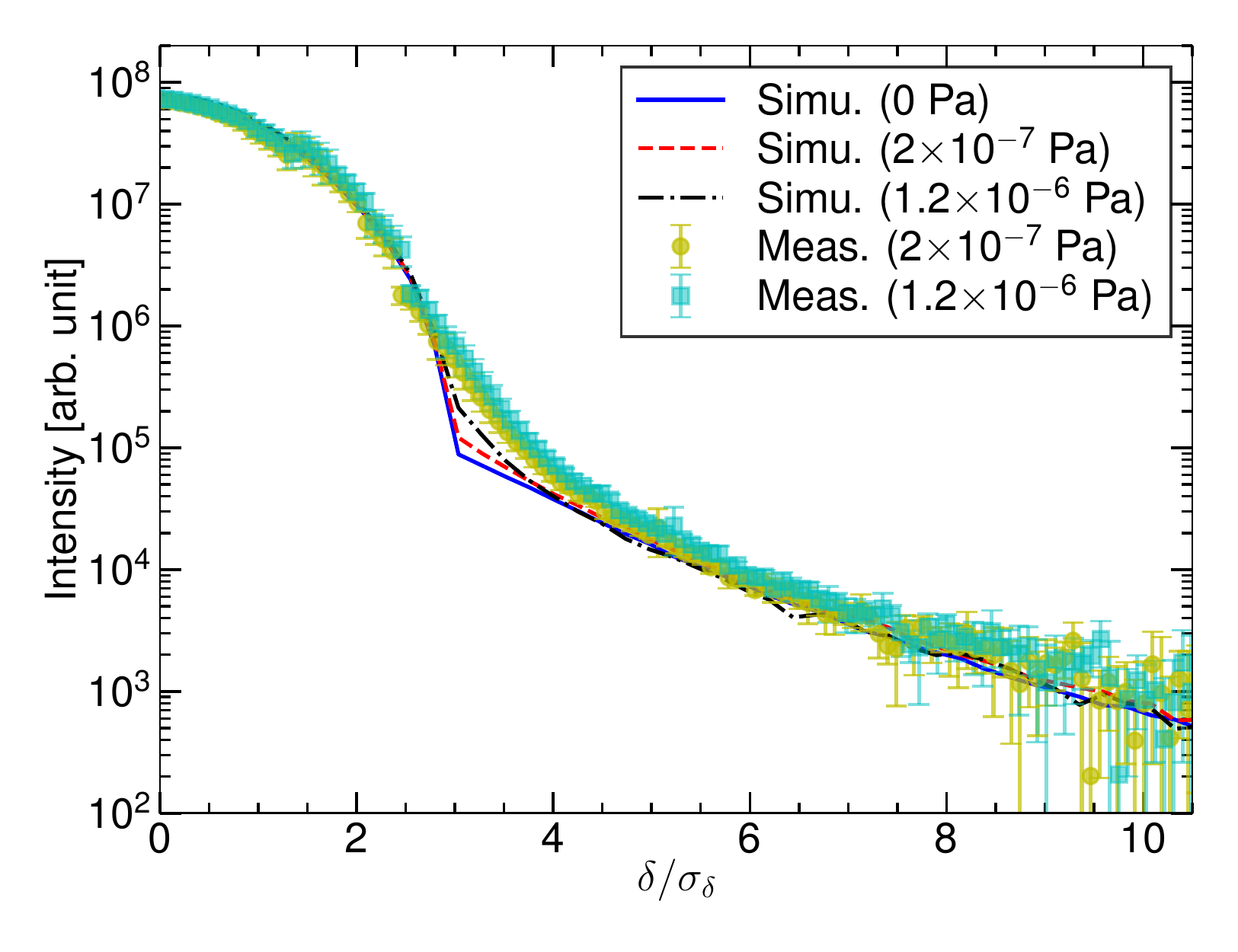}
    	\put(-200, 50){(c)}%\\    	
	    \includegraphics[width=0.49\linewidth]{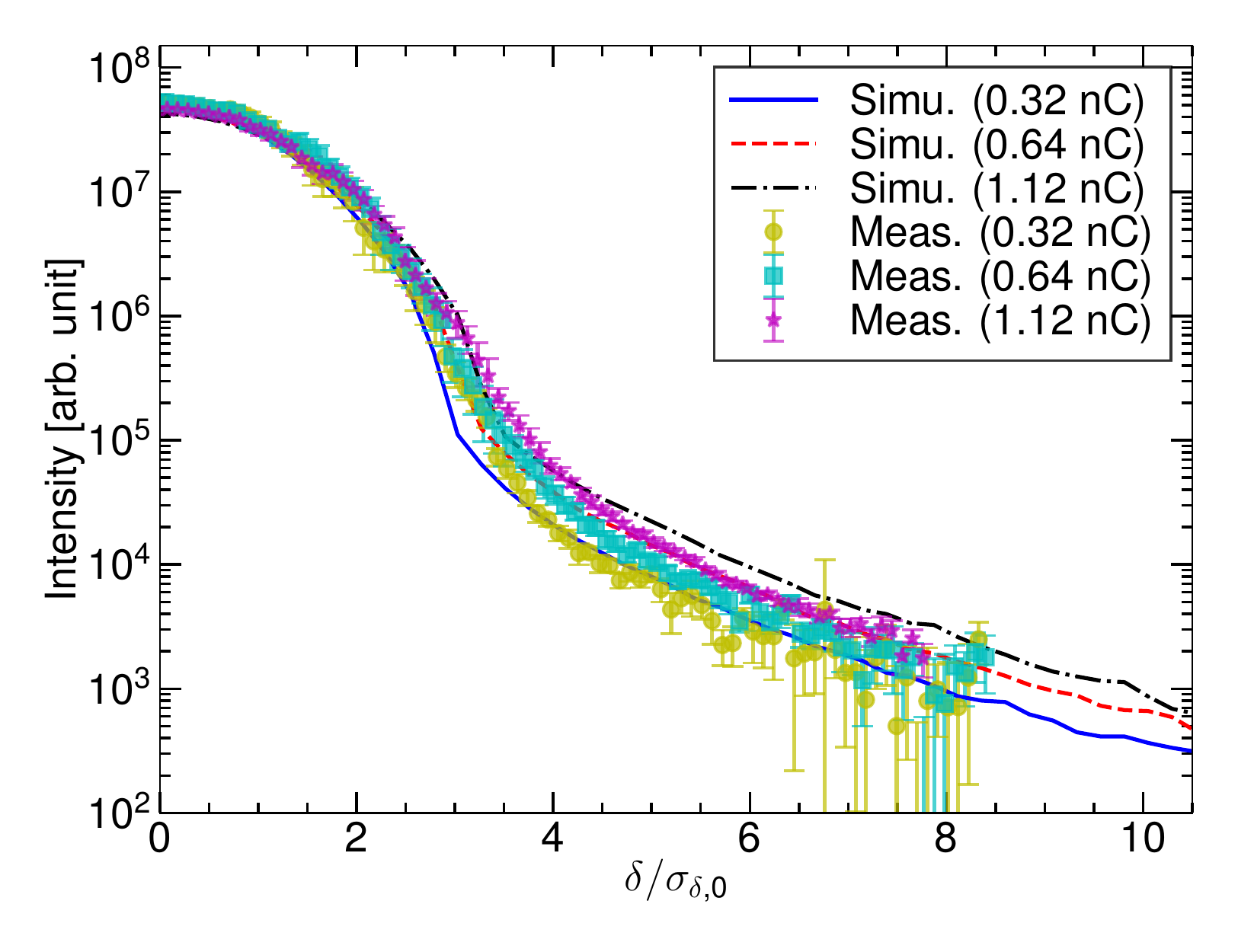}
    	\put(-200, 50){(d)}	    	
    	\caption{Normalized horizontal~(a, b) and momentum~(c, d) halos as a function of gas pressure and bunch charge. $\sigma_{x,0}$ represents horizontal beam size for a bunch charge of 0.48~nC and $\sigma_{\delta,0}$ is the energy spread at a bunch charge of 0.32~nC. Besides, a bunch charge of 0.48~nC is used for the vacuum-dependence studies. }
    	\label{fig:hor_dp_halos}
    \end{figure*}

The vertical halo was already shown in a previous study to be driven by the elastic BGS process~\cite{yang2018evaluation}, therefore only the horizontal and momentum halo are presented here. As shown in Fig.~\ref{fig:hor_dp_halos}~(a-b), the measured horizontal halos are in reasonable agreement with the simulations for different gas pressures~(2$\times10^{-7}$-1.2$\times10^{-6}$~Pa) and beam intensities~(0.16-0.96~nC). The horizontal halo is not influenced by the DR vacuum but is instead enhanced for a higher beam intensity. Due to the considerable background noise near the edges of the camera sensor, the measurements extend only to 8-10$\sigma$. One may see that the measurements away from the beam core are already noisy, especially for a low bunch charge.  
    
The momentum halos are imaged in the vertical plane with a vertical dispersion of about 200~mm at the monitor. The possible impact of the vertical betatron halo must therefore first be checked. The observations at two different gas pressures~(2$\times10^{7}$ and 1.2$\times10^{-6}$~Pa) show good agreement with numerical simulations and insignificant correlation with gas pressure, revealing a negligible contribution from the vertical betatron halo, as shown in Fig.~\ref{fig:hor_dp_halos}~(c). Notice that the momentum halo at 3-5$\sigma$ is somewhat underestimated due to the absence of multiple small-angle scattering process in simulation. For higher gas pressures, some increase in the predicted momentum halos in this region results from the inelastic BGS process. However, such an increment is not observed in the measurements. Although the observations show a weaker intensity dependence than the numerical predictions, the trends are consistent, as shown in Fig.~\ref{fig:hor_dp_halos}~(d). 
\begin{figure} %[h]
    \centering
	\includegraphics[width=0.95\linewidth]{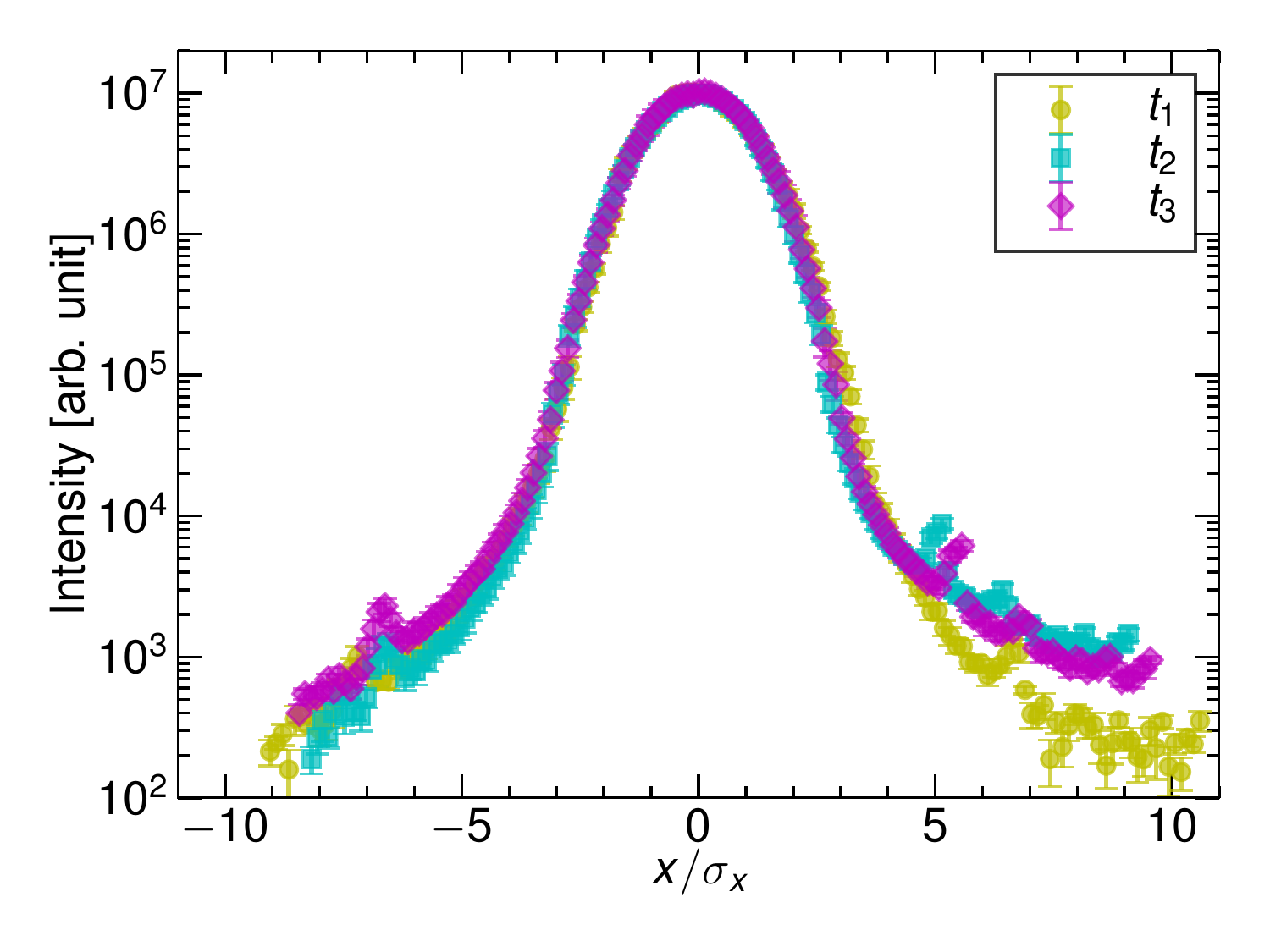}
    \caption{Horizontal halos versus extraction timing. The protuberances in the halo region arise from dust on the screen.}
    \label{fig:halo_x_kicker}
\end{figure}

\begin{figure} %[h]
    \centering
	%\includegraphics[width=0.95\linewidth]{Ntous_emityDep_v1}
	%\put(-50, 100){(a)}\\    	
	\includegraphics[width=0.95\linewidth]{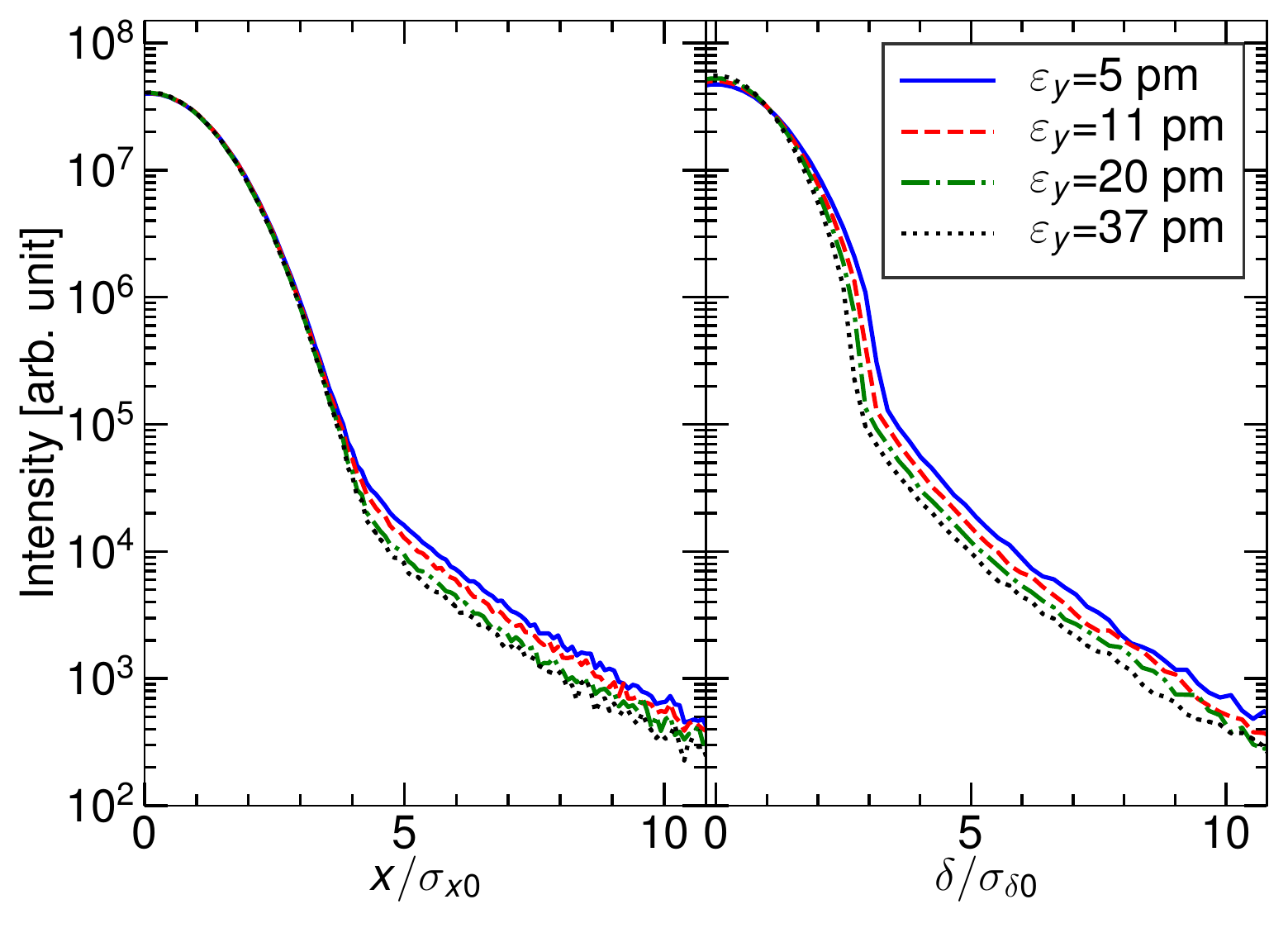}
	%\put(-195, 40){(b)}	    
	%\put(-90, 40){(c)}	    
    \caption{%Growth rate of Touschek events inducing an energy deviation of [0.2\%, 2\%]~(a), and 
    Horizontal and momentum halos versus vertical emittance with a bunch charge of 0.96~nC. $\sigma_{x0}$ and $\sigma_{\delta0}$ are for a vertical emittance of 10~pm.}
    \label{fig:halo_x_dp_emity}
\end{figure}

The reasonable general agreement between measurements and simulations, together with the dependencies consistent with expectations with respect to gas pressure and beam intensity, clearly points to Touschek scattering as the dominating mechanism forming horizontal and momentum halos. The residual discrepancies between predictions and observations might be the result of an imperfect extraction-kicker field or of inaccurate modeling of operational beam parameters. As shown in Fig.~\ref{fig:halo_x_kicker}, the measured horizontal halos can become asymmetric and slightly enhanced for inappropriate kicker timings. Optimization of the kicker timing is generally conducted before halo diagnostics, but the possible effects of residual timing errors might not be completely mitigated. The numerical predictions are also influenced by the uncertainties in vertical emittance measurements, errors in the mimicking of the realistic machine parameters, and ambiguities in the calculations of beam emittances and diffusion maps. A slight error on the model vertical emittance can for instance lead to notable differences in equilibrium beam sizes, Touschek scattering rate, and finally predicted horizontal and momentum halo distributions, as shown in Fig.~\ref{fig:halo_x_dp_emity}. An underestimated model vertical emittance for the numerical predictions may thus partly explain the discrepancy found between the observations and simulations at high beam intensity. More comprehensive measurements of emittances, bunch length and energy spread would therefore be needed as input to the simulations to improve the comparisons with the measurements. Moreover, halo distributions may be affected by resonances, chromaticity and nonlinearity, which should also be considered for future investigations.

\section{\label{sec:conclu} Conclusion}
The origin of the horizontal and momentum halos has been theoretically and experimentally studied for the KEK-ATF. A halo generator containing diffusion, BGS and Touschek scattering processes has been developed in a simulation approach based on realistic operational beam parameters. For halo diagnostics, a combined YAG/OTR monitor has been designed and installed in places where the dispersion can be adjusted for acquiring also the momentum-halo distribution. The reasonable consistency between observations and simulations for several gas pressures and beam intensities indicates that the Touschek scattering dominates both horizontal and momentum halos. Some observed residual discrepancies are attributed to an imperfect extraction-kicker field and remaining inaccuracies in the modeling of beam emittances. Further simultaneous measurements of emittances and beam halos employing an improved monitor with a higher dynamic range~{($\geqslant$$10^6$)} would be recommended.

The observations provide a reliable benchmark of the halo generator and validate its applicability to other GeV-scale low-emittance storage rings. Moreover, the importance of including multiple-scattering processes and proper measurement based accelerator modeling has been highlighted.

\section{Acknowledgements}
The authors would like to express their gratitude to the ATF2 collaboration and the staff of ATF. We also thank K. Oide, T. Lefèvre, S. Mazzoni, R. Nagaoka and D. Zhou for many helpful discussions and continuous encouragement. 
% This work was supported by the Toshiko Yuasa France-Japan Particle Physics Laboratory (project A-RD-10) and the MSCA-RISE E-JADE project, funded by the European Commission under grant number 645479.

\bibliography{aps_ref}

\end{document}